\documentclass[%
 aip,
 jmp,%
 amsmath,amssymb,
 reprint,%
]{revtex4-1}

\usepackage{graphicx}
\usepackage{dcolumn}
\usepackage{bm}
\usepackage{epstopdf}
\usepackage{amssymb}   
\usepackage[version=3]{mhchem} 
\usepackage{textcomp, libertine}
\usepackage{color}

\begin{document}

\preprint{AIP/123-QED}

\title[Allodi et al.]{Nonlinear TeraHertz Coherent Excitation of Vibrational Modes of Liquids}

\author{Marco A. Allodi}
\email[]{mallodi@uchicago.edu}
\altaffiliation{Current Address: Department of Chemistry, The Institute for Biophysical Dynamics, and The James Franck Institute, The University of Chicago, Chicago, IL, 60637}
\affiliation{Division of Chemistry and Chemical Engineering, California Institute of Technology, Pasadena, CA, 91125}

\author{Ian A. Finneran}
\affiliation{Division of Chemistry and Chemical Engineering, California Institute of Technology, Pasadena, CA, 91125}

\author{Geoffrey A. Blake}
\email{gab@gps.caltech.edu}
\affiliation{Division of Chemistry and Chemical Engineering, California Institute of Technology, Pasadena, CA, 91125}
\affiliation{Division of Geological and Planetary Sciences, California Institute of Technology, Pasadena, CA, 91125}

\begin{abstract}
We report the first coherent excitation of intramolecular vibrational modes via the nonlinear interaction of a TeraHertz (THz) light field with molecular liquids.
A TeraHertz-TeraHertz-Raman pulse sequence prepares the coherences with a broadband, high-energy, (sub)picosecond TeraHertz pulse, that are then measured in a TeraHertz Kerr effect spectrometer via phase-sensitive, heterodyne detection with an optical pulse. The spectrometer reported here has broader TeraHertz frequency coverage, and an increased sensitivity relative to previously reported TeraHertz Kerr effect experiments. Vibrational coherences are observed in liquid diiodomethane at 3.66~THz (122~cm$^{-1}$), and in carbon tetrachloride at 6.50~THz (217~cm$^{-1}$), in exact agreement with literature values of those intramolecular modes. This work opens the door to 2D spectroscopies, nonlinear in TeraHertz field, that can study the dynamics of condensed-phase molecular systems, as well as coherent control at TeraHertz frequencies.

Keywords: TeraHertz, Nonlinear and 2D Spectroscopy, Molecular Dynamics
\end{abstract}
 
\maketitle

\section{Introduction}
The microscopic structure and dynamics of condensed-phase matter determine its physical behavior. 
In liquids, the different types of interactions between a molecule and its surroundings fundamentally shape those dynamics and determine properties as diverse as vapor pressure, electrical conductivity, and a liquid's ability to solvate different compounds. Gaining new insight into the relationship between molecular dynamics and molecular interactions requires a continued expansion of the frontiers of light-matter interactions.

The TeraHertz (THz, 3-330~cm$^{-1}$, 0.1-10~THz), or far-infrared, region of the electromagnetic spectrum is one such frontier that offers novel probes of a variety of condensed-matter systems such as charge transport in semiconductors,\cite{Rice2013,Zaks2012} the biophysics of vision,\cite{Groma2008} and the coherent control of molecular reactions.\cite{LaRue2015} Of particular interest to physical chemistry, the THz-active degrees of freedom correspond to soft, large-amplitude modes of a liquid that participate directly in the molecular dynamics. At 298~K, $k_BT$ is 207~cm$^{-1}$. Thus, given the energy of THz modes relative to $k_BT$, these soft, predominantly intermolecular, modes are easily populated near room temperature, and require further study because measuring their dynamics may be key to understanding the properties of liquids such as water.\cite{SAHamm2013}

A central, open question about such molecular interactions involves the nature of molecular vibrations in liquids and how the vibrational dynamics on a femtosecond-to-picosecond (fs-to-ps, or ultrafast) timescale affect the physical properties of the fluid. The class of nonlinear, ultrafast, spectroscopic measurements, in the perturbative regime, that scale as the applied electric field cubed, generally known as third-order, or $\chi^{(3)}$, nonlinear spectroscopies, have a proven ability to interrogate (sub)ps dynamics. For isotropic systems, such as amorphous glasses or molecular liquids, the $\chi^{(3)}$ terms are the first nonlinear contributions to the perturbative expansion of the polarization due to the symmetry of the system.\cite{Boyd,Mukamel}  Specifically, two-dimensional (2D) optical and infrared techniques as well as Optical Kerr effect (OKE) spectroscopy are robust third-order methods that continue to enrich our understanding of liquids.\cite{Fayer2001} The third-order polarization measured in these experiments reports upon the beating of coherences that are generated and perturbed, in a controllable fashion, by the laser pulses delivered to the sample. 

However, for 2D measurements in the mid-IR, the modes studied are not thermally populated and thus only indirectly report on the dynamics of the THz-active degrees of freedom. Given the considerable challenges of simply generating and detecting intense THz pulses, the first attempt to investigate these soft THz modes of liquids via sub-ps 2D spectroscopy was with a 2D Raman-based approach.\cite{Tokmakoff1997} This technique never became broadly implemented because the measured signal is rigorously fifth-order in electric field ($\chi^{(5)}$), which leads to complicated alignment and weak signals that are overwhelmed by cascaded third-order processes.\cite{Blank1999,Jansen2000}

Developing a third-order 2D THz spectroscopy would enable new experiments to explore the structure and dynamics of condensed-phase systems. Indeed, pioneering 2D THz work has been done on a variety of systems such as semiconductor quantum wells,\cite{Kuehn2009,Kuehn2011} graphene,\cite{Woerner2013} and nonlinear crystals.\cite{Somma2014} These techniques demonstrate the breadth of science that can be addressed with THz techniques.

 Another approach to third-order THz spectroscopy of molecular systems has made use of {\em hybrid} pulse sequences, most successfully 2D Raman-THz.\cite{SAHamm2013} Such studies have shown that  Raman-THz-THz, THz-Raman-THz, and THz-THz-Raman pulse sequences can be quite sensitive to intermolecular modes because they combine both dipolar and polarizability interactions. The combination of interactions specifically provides insight into the anharmonicity of both the modes (mechanical) and the dipole moment surface(s).\cite{Ikeda2015,Hamm2012} These hybrid techniques are distinct from 2D Raman spectroscopy, a $\chi^{(5)}$ approach where the field interactions occur via the polarizability, and from OKE spectroscopy, which can measure linear Raman spectra and also employs only polarizability interactions.\cite{McMorrow1991,Righini1993} 
While the pulse sequences used in 2D Raman-THz spectroscopy (Raman-THz-THz, and THz-Raman-THz) enable investigations of molecular dynamics, the measured signals are linear in THz field.\cite{SAHamm2013}  

Nonlinear THz approaches would provide a means of manipulating coherences and exploring the dynamics of molecular systems directly via interactions on the thermally-populated potential energy surface --  offering the possibility of measuring and controlling the dynamics of a liquid by interacting with the modes that directly result from the intermolecular interactions. As discussed almost 20 years ago by Okumura and Tanimura, tools such as 2D THz spectroscopy would complement Raman-based approaches that required the involvement of high-lying virtual states, and provide new routes to the observation of dynamics that may not be visible as perturbations to a mode thousands of wavenumbers above the energy of the intermolecular modes (e.g. an intramolecular OH stretch around 2500~cm$^{-1}$ to 3500~cm$^{-1}$).\cite{Okumura1998}

A THz-THz-Raman (TTR) pulse sequence, for example, would augment those already used in 2D Raman-THz spectroscopy\cite{Ikeda2015} and generate a $\chi^{(3)}$ signal, which is nonlinear in THz field. The TTR pulse sequence is identical to THz Kerr Effect (TKE) spectroscopy, where large THz field strengths (in excess of 100 kV/cm) create a transient birefringence that is detected by the optical pulse, making TKE spectroscopy the first technique to measure a signal that is nonlinear in THz field on molecular systems.\cite{Hoffmann2009}  Additionally, in a centrosymetric material, this pulse sequence yields a backgroundless measurement of a $\chi^{(3)}$ process. It is the backgroundless/low-background nature of many 2D IR measurements that has been a major part of the success of that technique.

Here, we report the first nonlinear coherent excitation of intramolecular vibrational modes of liquids using THz Kerr effect spectroscopy. We employ a TTR pulse sequence to initiate intramolecular vibrational coherences via the nonlinear interaction of a THz light field with the liquid. With this experimental setup, the excitation proceeds via a resonant-two-photon-like mechanism, which is a distinct from a Raman excitation pathway. While coherent excitation of THz molecular modes was first observed over 25 years ago in the gas phase,\cite{vanExter1989} the measurement of nonlinear coherent excitation in room temperature liquids opens the door to 2D spectroscopies, which are nonlinear in THz field, and that could ultimately allow for the coherent control of low-frequency vibrational modes in the condensed phase.

\section{Experimental Methods}
 \begin{figure}[ht]
 \includegraphics[width=8.8cm]{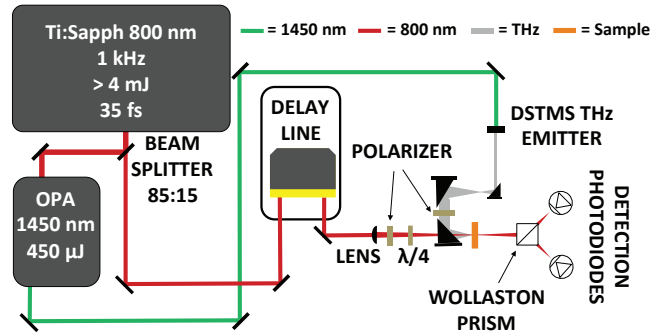}%
 \caption{ The heterodyne-detected THz Kerr effect spectrometer used for THz-THz-Raman experiments. Note that the 800~nm probe beam is {\it s}-polarized and that the THz polarizer is set at 45$^{\circ}$ with respect to that polarization. \label{spectrometer}}%
 \end{figure}

 A diagram of the spectrometer used in these experiments can be seen in Figure~\ref{spectrometer}. The heart of the laser system is a Legend Elite USP Ti:Sapphire regenerative amplifier (Coherent, Inc.) that produces 35~fs, 4.5~mJ pulses, centered around 800~nm, at a 1~kHz repetition rate. These pulse are used to drive an ultrafast optical parametric amplifier (OPA, Light Conversion Ltd.) that down-converts the pulses further into the near-IR. Approximately 450~mW of 1450~nm light from the signal beam of the  OPA passes through a 4-N,N-dimethylamino-4$^{\prime}$-N$^{\prime}$-methyl-stilbazolium 2,4,6-trimethylbenzenesulfonate (DSTMS) crystal (Rainbow Photonics AG) set on a 3~mm diameter aperture to produce $\sim$160~nJ THz pulses. 
After magnifying the beam waist by a factor of 7.5, the THz beam passes through a wire-grid polarizer set at 45$^{\circ}$ relative to the polarization of the 800~nm probe beam and is then focused onto the sample to generate a THz electric field in excess of 300~kV/cm.  To maximize the magnitude of the THz electric field at the sample, the chirp of the 800 nm regenerative amplifier pulses is adjusted while monitoring the THz generation efficiency via electro-optic sampling.\cite{Wu1995,Lu1997}

Roughly 2~mW of the regenerative amplifier 800~nm light  is split off before the OPA and sent down a delay line (Newport ILS150HA with ESP300 controller) fitted with a retroreflector for precise control of the arrival time of the 800~nm probe pulse at the sample. The probe beam is focused through a hole in the back of the THz-focusing off-axis parabolic mirror in a co-linear geometry with the THz pulse. Both beams hits the sample that is held in a Suprasil quartz (fused silica) cuvette of 1~mm path length.

The 10$^5$:1 polarizer  (Throlabs LPVIS050-MP2), $\lambda$/4 plate, and a Wollaston prism placed after the sample -- that serves to split the orthogonal polarizations of the probe beam -- are critical to the heterodyne-detection scheme. By adjusting the 800 nm polarizer such that almost all the light is in the {\it s}-polarization, the small fraction of light that is {\it p}-polarized acts as a 90$^{\circ}$ out-of-phase local oscillator that is heterodyned with the molecular Kerr effect signal on a pair of photodiodes. This scheme ensures that only the backgroundless birefringence signal is measured for centrosymmetric media.\cite{McMorrow1991,Mukamel} An iris in front of the photodiode is used to attenuate the {\it s}-polarized beam so that it is of a comparable magnitude to the {\it p}-polarized beam.\cite{Brunner2014,Johnson2014} The 1450 nm beam is modulated with an optical chopper at 500~Hz and photodiode signals are sent to a digital lock-in amplifier (SRS SR830), where they are subtracted, and the TTR signal is measured.  We verified that the detected TTR signal scales as the applied THz electric field squared ($E_{THz}^2$) by inserting high-resistivity Si plates into the THz path before the wire-grid polarizer and measuring the peak TTR signal of carbon disulfide (Sigma-Aldrich, $>$99\%). A plot of this data can be seen in the supplemental information, Figure S1.\footnote{See supplemental material at [URL will be inserted by AIP] for the plot of TTR signal scaling as a function of applied THz field, Figure S1, a calculation of THz field strength, and the frequency-domain data out to the Nyquist bandwidth of the measurement, Figure S2.} More details regarding the instrumentation and optics can be found in our earlier work on linear THz spectroscopy and references therein.\cite{Allodi2014,Ioppolo2014}

The liquids studied in this experiment were $>$99\% purity and were used as received without further purification. All measurements were taken in a nitrogen (\ce{N2}) purged atmosphere to remove the components of ambient air, especially water, that are strong THz absorbers. The optical quality of the fused silica cuvette and liquid samples were such that minimal scattered 800~nm light was observed. This significantly increases the sensitivity of the TKE spectrometer.
\vspace{1.8cm}
\section{Results and Discussion}
\vspace{.3cm}
The third-order electric field measured, $E^{(3)}_{Sig}(t)$, is directly proportional to the third-order molecular polarization $P_{i}^{(3)}(t)$ after a phase shift. The polarization is given by:
\begin{equation}
P_{i}^{(3)}(t,\tau)=E^{pr}_j (t-\tau) \int^{\infty}_0 d\tau ^{\prime} \, R^{(3)}_{ijkl}(\tau^{\prime})E^{THz^*}_k (t-\tau^{\prime})E^{THz}_l (t-\tau^{\prime}), \:\:\:
\end{equation}
where $R^{(3)}_{ijkl}(\tau)$ is the 3$^{rd}$-order response function of the liquid, $\tau$ is the time between the THz and 800~nm pulses, and $E^{pr}(t-\tau)$ is the probe pulse electric field.\cite{Zanni,Mukamel}
The response function, $R^{(3)}_{ijkl}(\tau)$,  where the indices correspond to polarization, can be separated into electronic and nuclear components. Since the THz field is non-resonant with any electronic transitions, the electronic component will only contribute to the signal when the applied THz electric field is present, and only the nuclear components of the response function contain information about the molecular dynamics. Given that $R^{(3)}_{ijkl}(\tau)$ is directly related to the emitted polarization, we can relate the measured signal, $S(\tau)$, to the molecular dynamics initiated with the THz pulse.

 \begin{figure}[ht]
 \includegraphics[width=8.6cm]{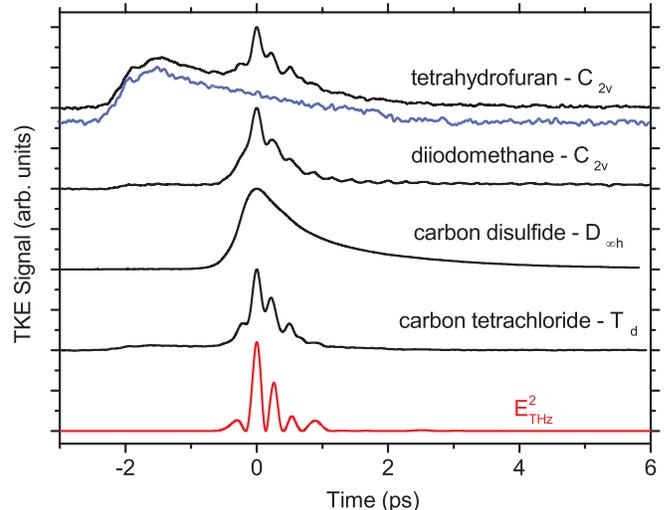}%
\vspace{-.5cm}
 \caption{Measured third-order HD-TKE signals, generated by a TeraHertz-TeraHertz-Raman pulse sequence, of the liquids in this study, labeled with the point groups of the isolated molecule. Vibrational coherences are clearly visible as oscillations in the diiodomethane and carbon tetrachloride signal. Carbon disulfide yielded the largest response. The bottom trace is the square of the THz electric field applied in these measurements.  At time zero, the 800~nm pulse is probing the peak TKE response of the liquid. Positive times on the abscissa denote an interaction with the THz pulse applied first followed by the 800~nm detection pulse. The blue trace below the tetrahydrofuran data shows the empty cuvette background response. Note that all molecular data are normalized and vertically offset for clarity.\label{TD_data}}%
 \end{figure}

 The heterodyne detection scheme employed allows for the maximum recovery of information about $E^{(3)}_{Sig}(\tau)$. Since the photodiodes that detect the 800~nm probe light function as square-law detectors, the intensity of the light ($I(\tau)$) is measured and the combined signal is integrated over the time variable $t$ because the detectors are not fast enough to measure the oscillating signal in time directly. Without a local oscillator provided by another light field, the homodyne signal can be written as:
\begin{equation}
\begin{centering}
S_{hom}(\tau)=\int^{\infty}_{0}dt \:|E^{(3)}_{Sig}(t,\tau)|^2 = I_{sig}(\tau). 
\end{centering}
\end{equation}
By taking the square of the signal and integrating over time, information about the phase of the light field is lost. This phase information is especially important if the signal contains any oscillatory components going as $e^{i\omega t}$, since the integral over the modulus squared would cancel these oscillatory terms.\cite{Zanni}

 \begin{figure}[ht]
 \includegraphics[width=9cm]{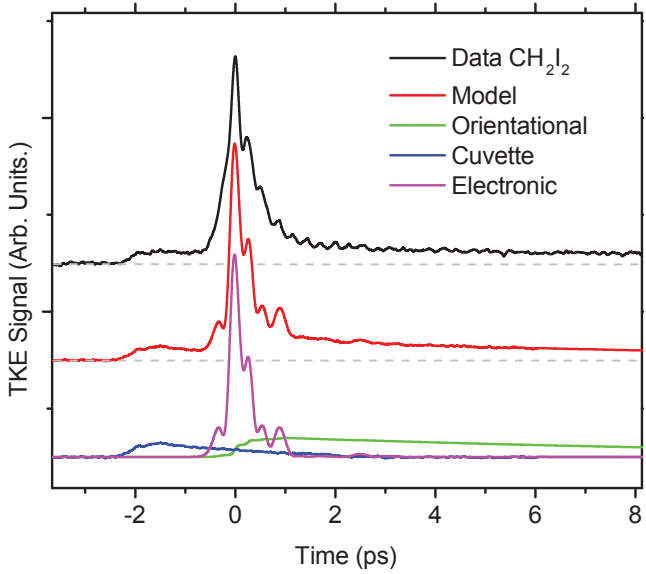}%
 \vspace{-0.1cm}
 \caption{ Modeling of the HD-TKE signal. The experimental data for \ce{CH2I2} are plotted at the top in black. The complete model is plotted directly below in red. The individual components of the fit, cuvette (blue), orientational (green), and electronic (magenta), are plotted together on the final line.\label{model}}%
 \end{figure}
 
 In contrast, the heterodyne-detected signal can be written as,
\begin{eqnarray}
S_{het}(\tau)=\int^{\infty}_{0} dt \: |E_{LO}(t-\tau) + E^{(3)}_{Sig}(t,\tau)|^2\: \: \:\:\: \\
=  I_{LO}(\tau) + I_{sig}(\tau) + 2 Re(\int^{\infty}_{0} dt  E_{LO}(t-\tau) E^{(3)}_{Sig}(t,\tau)),\:\: \:\:\:
\end{eqnarray}
 where $Re$ denotes the real part of a complex number that can be measured by physical detectors. The contribution from $I_{LO}(\tau)$ can be removed by a number of common-mode-rejection techniques. In this work, we remove $I_{LO}(\tau)$ by using balanced photodiodes as discussed above. The $I_{sig}(\tau)$ term can be neglected since is small compared to the LO and signal interference term. Thus, heterodyne detection increases the instrumental sensitivity since the measurement scales both with the magnitude of the local oscillator electric field, created using the polarizer, and that of the molecular signal's electric field. We confirmed the linearity of the heterodyne detection by adjusting the power of the 800~nm beam and observing that the measured signal scaled linearly with the applied 800~nm average power. 

 If the duration of the LO is shorter than the timescale of the dynamics of interest in $R^{(3)}_{ijkl}(\tau)$, then the envelope of LO pulse can be thought of as a sampling function and approximated as a delta function.\cite{Torre2007}  As such we can write:
\begin{equation}
S_{het}(\tau) = 2 Re(\int^{\infty}_{0} dt \, \delta(t-\tau) E^{(3)}_{Sig}(t,\tau)) = 2 Re( E^{(3)}_{Sig}(\tau)).
\end{equation}
  This makes heterodyne detection sensitive to the phase of the emitted signal field, as opposed to homodyne measurements that only capture the signal-field squared. In reality, it is the phase difference that survives the integral, but since the phase of the LO is known, the phase of the signal is also known. The relationship described in eq. (3) ensures that the heterodyne signal is linearly proportional to the response function while the homodyne signal is proportional to the response function squared.\cite{Torre2007}

 The signals generated by a TTR pulse sequence in a variety of room temperature liquids, along with the trace of the THz electric field squared ($E^*_{THz}E_{THz}$), measured via electro-optic sampling in a GaP(110) crystal \cite{Wu1995,Lu1997}, are presented in Figure~\ref{TD_data}. We refer to these equivalently as heterodyne-detected TKE (HD-TKE) or TTR signals. The zero point on the time axis corresponds to the point where the peak of the 800~nm pulse is overlapped with the peak of the THz pulse, yielding maximum signal. At positive times, the 800~nm pulse interacts with the sample after the peak of the THz pulse, while at negative times, the 800~nm pulse interacts with the sample first. This time delay is controlled to high accuracy and precision with an opto-mechanical delay line. 

 The measured nonlinear TTR response near zero time follows the electric field squared of the THz pulse (E$^2_{THz}$).\cite{Zhong2008} 
The nuclear part of the signal remains zero until $\tau=0$ and contains all of the potentially retrievable information about the molecular dynamics of the liquid.\cite{Righini1993} It is well established in the literature that rotational diffusion of individual monomers returning to an isotropic distribution of molecular orientations is the dominant contribution to the OKE and TKE signals starting several picoseconds after the peak, and that this post-input pulse response can be fit with a decaying exponential. For example, such behavior is clearly visible in the carbon disulfide (CS$_2$) HD-TKE data presented in Figure~\ref{TD_data}. The large molecular polarizability of CS$_2$ leads to a large electronic response followed by an orientational response that quickly dominates the signal.

 The recovery of a nonlinear TTR signal in tetrahydrofuran (THF, EMD Millipore, $>$99.5\%) demonstrates the significant sensitivity improvements that result from heterodyne detection. Our data show a clear response from THF, which was not seen in previous work. The pre-pulse seen in the data before 0~ps is the TTR signature of the fused silica sample cuvette. It is present in all of the data in Figure~\ref{TD_data}, and provides a reference for the relative molecular responses.  The blue trace in Figure~\ref{TD_data} shows the cuvette response, scaled to match the pre-pulse seen in the the THF trace and slightly offset for clarity. The cuvette signal lasts for just over 4~ps. It is quite similar in size to that of the tetrahydrofuran signal, but stable from experiment to experiment. As a result, the cuvette response can be removed from the full trace to isolate the nonlinear TTR signal of strongly (THz) absorbing liquids. Straight-forward improvements to the apparatus, such as the inclusion of a liquid jet, will improve the sensitivity in future experiments. Finally, we note that the cuvette signal is so weak relative to the TKE response of the other liquids studied that its presence in the data does not affect the conclusions drawn from these experiments.
 
 To further understand the interaction of the THz pulse with diiodomethane (Sigma-Aldrich, $>$99\%), we used a three component model of the data, including electronic, orientational, and cuvette responses, the components and results of which are seen in Figure~\ref{model}. For the electronic response we applied a minimal moving average to the E$^2_{THz}$ pulse measured by electro-optic sampling. The orientational response was modeled based on the previous work of Hoffman et al.\cite{Hoffmann2009} by stepping over the measured E$^2$ pulse with the following finite-difference solution to the differential equation:
\begin{equation}
S(t+ \Delta t)= S(t) + [E^2(t+ \Delta t) - S(t)/\tau_o]*\Delta t ~~,
\end{equation}
where $S(t)$ is the measured birefringence signal, $\Delta t$ is the time step, $E^2$ is the measured THz pulse, and $\tau_o$ is the orientational decay time (13~ps for \ce{CH2I2}). The cuvette response was measured with an empty cuvette, and the full modeled response is a linear combination of the three components that are each scaled to match the data. The qualitative agreement with the results is quite good. When all three components are considered, a fourth oscillatory signal is clearly visible in the CH$_2$I$_2$ and CCl$_4$ traces.  

 \begin{figure*}[ht]
 \includegraphics[width=\textwidth]{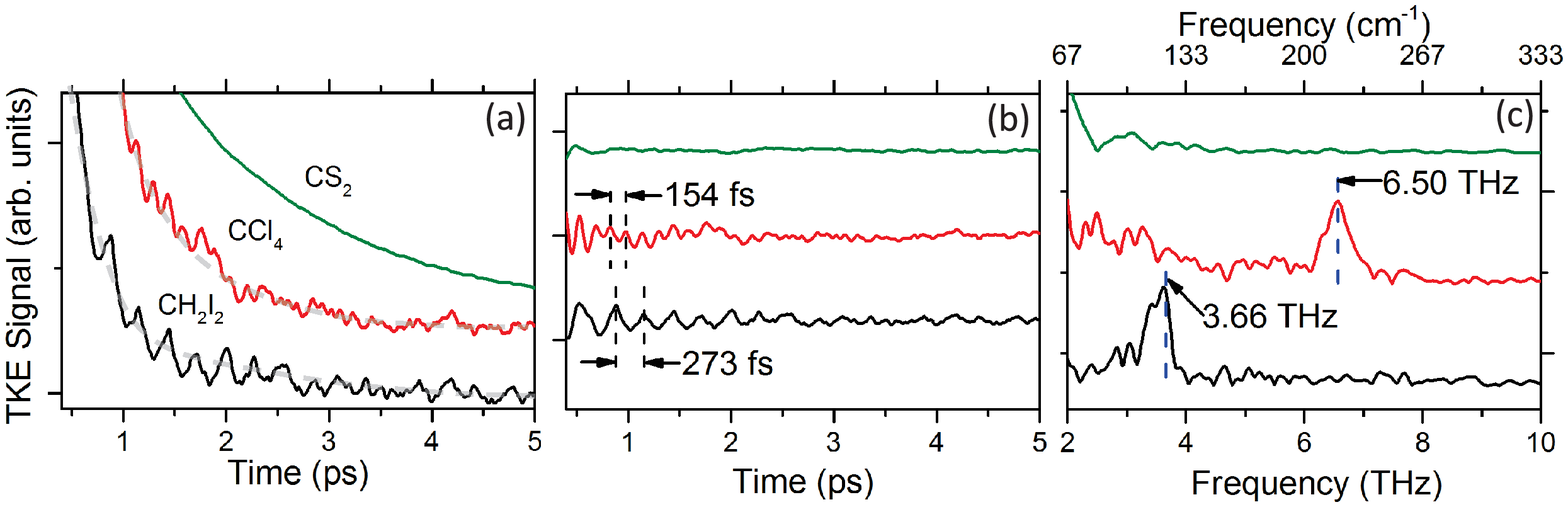}%
 \vspace{-0.4cm}
 \caption{a) Vibrational coherences in the diiodomethane (CH$_2$I$_2$) and carbon tetrachloride (CCl$_4$) TKE signals. Carbon disulfide is plotted for reference. b) The detrended TKE signals for CH$_2$I$_2$, and CCl$_4$. A biexponential fit was used to remove the contribution of the orientational response on the signal. Dashed black lines indicate the period of the coherences observed in the two molecules. c) The magnitude of the complex Fourier transform of the CH$_2$I$_2$ and CCl$_4$ data. The dashed blue lines indicate the literature values of the experimentally measured vibrational mode frequencies. \label{beats}}%
 \end{figure*}
 
In diiodomethane, these previously unmeasured oscillations, highlighted in Figure~\ref{beats}a, correspond to a vibrational coherence that lasts for several ps, excited by the high-field-strength THz pulse. While the THz excitation pulse extends to roughly 1~ps after $\tau=0$ (as seen in the E$^2_{THz}$ trace at the bottom of Figure~\ref{TD_data}), the oscillating signal lasts for \textit{at least} another 5~ps. Both the electrons and nuclei in the molecule will respond to the THz field, but the Born-Oppenheimer approximation tells us that the electron motion will cease only a few fs after the THz pulse has propagated through the sample.\cite{Born1927} As a result, once the THz pulse no longer interacts with the sample, the only contribution to the THz Kerr effect signal arises from the nuclear portion of the response function. This separation of nuclear and electronic response has been demonstrated in the THz Kerr effect work of Hoffmann et al.\cite{Hoffmann2009}; is well established in the OKE literature \cite{Zhong2008}; and can be seen in the model of the data employed in Figure~\ref{model}.

Given that the coherence lasts much longer than the THz pulse, the observed oscillations cannot result from the electronic response to the applied THz field, but rather must arise from the coherently-excited molecular motions in the liquid. While the orientational component is also part of the nuclear response, the coherences remain after biexponential fits are used to remove both the initial electronic response, and the orientational response, as Figure~\ref{beats}b shows.Two distinct time constants in the orientational decay require a biexponential model for a good fit, with the shorter timescale resulting from molecular dynamics akin to those revealed in the so-called intermediate response in OKE spectroscopy. The use of the biexponential model and the intermediate response will be discussed in more detail in an upcoming publication.

These coherences were likely not observed in the previous TKE measurements on molecular systems because of the limited frequency content available via room-temperature LiNbO$_3$ tilted-pulse-front THz generation and because of the phase information, high sensitivity, and dynamic range provided by heterodyne detection. Finally, we checked the signal scaling of these oscillations and they match the scaling of the orientational response to the applied fields.

The nature of the chemical bonds in CH$_2$I$_2$ leads to a fundamental normal vibrational mode, $\nu_4$, at 3.66~THz (122~cm$^{-1}$). \cite{Johnson2006} We confirm that the coherence corresponds to this mode by taking a numerical Fourier transform of the data starting at time zero and using an asymmetric Hann window for apodization.\cite{Ioppolo2014,Galvao2007} The magnitude of the complex-valued Fourier transform (also known as the absolute value of the Fourier transform) is plotted in the right panel of Figure~\ref{beats}. Similar coherences are also observed in carbon tetrachloride (CCl$_4$, Sigma-Aldrich, $>$99.5\%). Here, the two interactions with the THz field are able to excite a coherence in the intramolecular $\nu_2$ mode at 6.5~THz (217~cm$^{-1}$)\cite{Gabelnick1976} that is also visible in the right panel of Figure~\ref{beats}. 

 These results are in excellent agreement with the literature values reported for the pure liquids\cite{Johnson2006,Gabelnick1976}, and well within the Fourier transform-limited spectral resolution of $\sim$5~cm$^{-1}$ produced by the 6~ps traces. Indeed, the difference between the literature value for the peak center and our measurement for diiodomethane is only 1~cm$^{-1}$. We also note that while the linewidth of these two features is produced by the observed dephasing time of the coherences, conclusive statements as to whether the homogeneous linewidth is commensurate with that observed in linear spectroscopy of the liquid(s) will require higher signal-to-noise data.

\begin{figure}[ht]
\includegraphics[width=8.8cm]{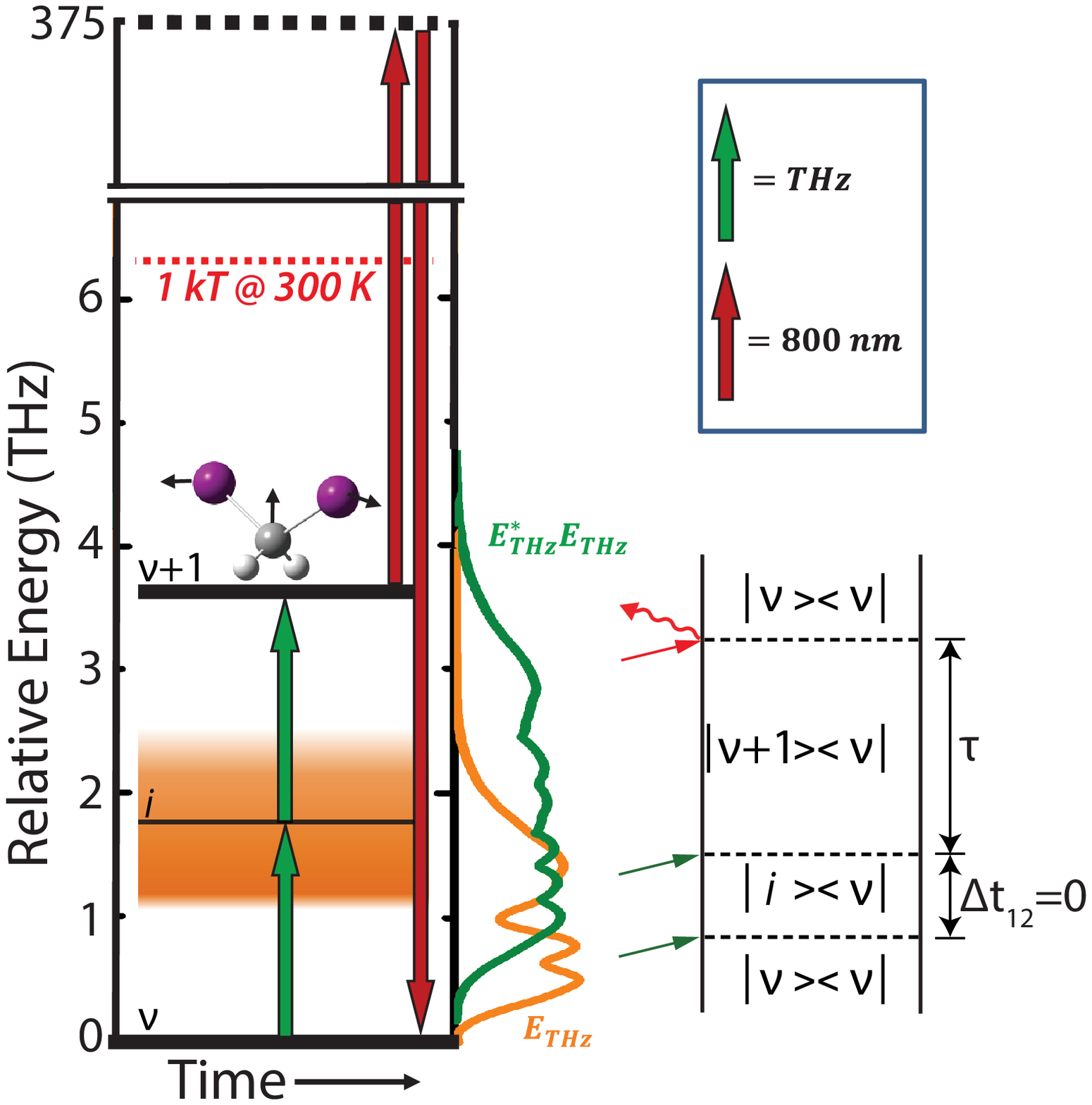}%
\vspace{-0.3cm}
\caption{A Jablonski diagram showing the relative energy of the interactions and a representative double-sided Feynman diagram showing one possible pathway that leads to the vibrational coherence of the $\nu _4$ vibrational mode in diiodomethane. The green arrows correspond to THz absorption processes. The red arrows indicate an anti-Stokes Raman process. Note that the two green arrows in the Feynman diagram are slightly separated by time $\Delta {\rm t _ {12}}$ for diagramatic purposes, however both interactions come from the same THz pulse as indicated by $\Delta {\rm t _ {12}}=0$. The orange trace on the right of the Jablonski diagram corresponds to the power spectrum of the THz electric field and shows the bandwidth covered by the pulse. The green trace corresponds to the power spectrum of the THz electric field squared. The spacing between THz and optical pulses is labeled $\tau$. These figures are drawn to show that the system starts with some number of vibrational quanta, $\nu$, and is excited through some intermediate state, $i$, to a coherence between a state of $\nu + 1$ and $\nu$, which generates the observed TKE signal. \label{Jablonski}}
\end{figure}

The results from diiodomethane and carbon tetrachloride are the first measurements of vibrational coherences that result from the \textit{nonlinear} coherent excitation with THz radiation ($E^2_{THz}$) in a condensed-phase, molecular system. The measured signal must be third-order in electric field (a $\chi^{(3)}$ process) given the isotropic nature of the samples. It is also backgroundless because we do not spectroscopically detect the linear THz response. Our signal is detected on Si photodiodes at 800~nm, which do not respond to incident THz photons, and we do not spectroscopically resolve the 800~nm photons we detect. Thus, we are not measuring a linear optical spectrum with our birefringence measurements. Finally, the second-order response is zero since $\chi^{(2)}$ is zero by symmetry in any isotropic system.\cite{Boyd} TKE measurements then detect the transient birefringence that results from the nonlinear coherent excitation of THz active modes. 

One can see that this is a birefringence measurement by considering the polarization of the THz excitation and optical probe beams. The third-order response function, $R^{(3)}_{ijkl}(\tau)$, is a rank 4 tensor, where the indices correspond to the polarization of the electric fields of the light applied. All $\chi^{(3)}$ nonlinear spectroscopies follow the same polarization rules. In the dipole approximation, the different polarizations must be in pairs, e.g. YYXX, since an odd number of terms, e.g. YYYX, would result in an odd number of cosine terms in the integral over the angular dependence of the response function, which would integrate to zero.\cite{Zanni} Thus, the allowed polarizations in the dipole approximation are XXXX, YYXX, XYXY, XYYX, (and further substitutions of X and Y), where, for example, XXYY would be pump with Y-polarization twice, pump on X polarization once, then detect emission from the X-polarization. Both OKE and TKE signals are a linear combination of XXXX and YYXX, since the probe pulse is tilted at 45 degrees with respect to the pump. Specifically, the pump is oriented in the X+Y polarization (+45 degrees), the 800~nm probe is aligned to be initially s-polarized, Y, and we detect a signal that is p-polarized, X. As shown by Mukamel, the fact that the signal is 90$^{\circ}$ out-of-phase from the polarization of the probe ensures that this samples the birefringence, as opposed to dichroism.\cite{Mukamel} In addition, the overall signal is XXXX-YYXX, equal to the XYXY response by symmetry,\cite{Torre2007} so R$^{(3)}_{ijkl}$ = R$_{XYXY}$ in eq. (1), which is the same polarization response as OKE spectroscopy. That is, the transient birefringence, which is analyzed with a crossed polarizer to yield the signal, is measured from the differences in polarized emission from the sample.

Thus, the coherences observed in these measurements are generated via two, simultaneous, THz field interactions analogous to resonant two-photon absorption. Figure~\ref{Jablonski} contains both a Jablonski diagram and a double-sided Feynman diagram illustrating this process. One can think of this as an absorption mechanism in a three level system, where two photons are absorbed during the temporal duration of the THz pulse. The system starts out with some number of vibrational quanta, $\nu$, on a given vibrational manifold. In one possible pathway, the first photon (corresponding to the first green arrow in either the Jablonski or Feynman diagram in Figure~\ref{Jablonski}) takes the system into a coherence between the $\nu$ state and some intermediate state, while the second photon excites the system further to generate the coherence between the $\nu$ and the $\nu +1$ state. Other pathways starting from excited vibrational states likely contribute, given that $k_BT$ at 300~K is 6.2~THz. However, in this work, the time between the two THz field interactions, denoted by $\Delta {\rm t _ {12}}$ in Figure~\ref{Jablonski}, cannot be independently controlled, and without this experimental handle, these measurements cannot determine the exact pathways. A mechanism involving a three-level system certainly would generate the oscillatory features observed in the TKE traces, and the Feynman diagram in Figure~\ref{Jablonski} represents one such pathway that would create such a coherence. 

Several different groups have used a three-level system to explain the nonlinear spectra of low-frequency modes in molecular systems. This is similar to the mechanism used by Fleischer~et~al.~to understand the two-quantum coherences generated by intense THz pulses in a gas-phase molecular sample,\cite{Fleischer2012} by Tokmakoff~et~al. to describe the pathways that contribute to 2D Raman signals,\cite{Tokmakoff1997} and by Hamm and Savolainen to describe hybrid Raman-THz spectroscopy.\cite{Hamm2012} Also, several authors have discussed how, for this type of mechanism, the signal should be sensitive to the anharmonicity of the vibrational modes, and that the effect of the anharmonicity should be quite clear in a 2D TTR experiment.\cite{Hamm2012,Ikeda2015} 

The orange trace on the right side of the Jablonski diagram in Figure~\ref{Jablonski} depicts the bandwidth of the THz pulse, which extends to roughly 2~THz. This is insufficient to directly excite the 3.66~THz mode of CH$_2$I$_2$. Further, we note that the Suprasil quartz cuvette used here has a strong cutoff starting around 3~THz, where $\alpha$ is around 15~cm$^{-1}$, increasing to $\sim$20~cm$^{-1}$ at 3.6~THz, and rising rapidly to $\alpha = 50~$cm$^{-1}$ by 5.5~THz.\cite{CunninghamPhD2010} As such, few 3.66~THz photons access the liquid, likely precluding a resonant dipolar mechanism involving any frequencies at or above the $\nu _4$ fundamental. 

The THz pulse energies available here place the experiment squarely in the perturbative regime.
As such, only third-order terms contribute to the observed signal, as can be shown by doing a simple estimation of the Rabi frequency for a 0.1~D dipole (a likely overestimation of the transition dipole of the 122~cm$^{-1}$ mode in CH$_2$I$_2$) in a 300~kV/cm electric field. When multiplied by the 200 fs FWHM of the THz field, a pulse strength of $\pi$/165 results. Excitation squarely in the strong-field regime requires a $\pi$/2 pulse, that would need THz-pulse field strengths of $\sim$30~MV/cm (or roughly 100 times the current field strength).

A strong-field excitation mechanism would lead to higher-order polarization terms contributing to the signal. If this were the case, one might expect to observe coherences between higher-lying vibrational modes as well, given our detection bandwidth. As discussed above, the coherences are detected in the time domain by adjusting the timing ($\tau$) between the arrival of the THz pulse and the $\sim$35~fs 800~nm pulse by sweeping a delay line. The delay line runs at a constant velocity, collecting points that amount to a 5~fs spacing. The convolution of the 35~fs pulse with this oversampling should yield an effective spacing of independent points in the time domain of 10-20 fs. The duration of the 800~nm pulse then limits the bandwidth of the measurement to 25-50 THz. As a result, if our generation mechanism did indeed generate coherences of higher-lying modes, up to $\sim$50~THz, they would be present in the data. We have confirmed that there are no such features in the data and have added a plot to the supplementary information showing the extended frequency-domain response.$^{27}$

Consequently, the only way to put more than 3~THz of energy into the vibrational mode of CH$_2$I$_2$ is via a process like resonant two-photon absorption involving a three-level system (shown in Figure~\ref{Jablonski}), which is clearly distinct from a Raman process. Given the 2~THz bandwidth contained within the pulse, two THz field interactions can potentially access a mode 4~THz above the initial energy of the molecule(s). However, the measured intramolecular mode of carbon tetrachloride sits at 6.5~THz. As such, the exact mechanism of excitation in CCl$_4$ is unclear. It may be a result of excitation from thermally excited states, and future experiments with a temperature controlled cell would help support or refute this hypothesis.

\section{Conclusion}
In this work, we have measured the coherent excitation of intramolecular vibrational modes initiated by a nonlinear TeraHertz interaction with a molecular liquid. While the details of the excitation mechanism are still unclear, two quanta of THz light are needed to generate the observed response, and the intermolecular modes below 3~THz may well be involved in enhancing the signal strength.\cite{Mukamel} One means of testing this conjecture is with studies of the temperature dependence of the TTR response. In liquids such as CH$_2$I$_2$, and CCl$_4$, THz frequency modes are thermally populated, and it is precisely these modes that determine the molecular dynamics. Thus, they are distinct from modes in the mid- or near-IR, where $h\nu \gg k_B T$.

As a result, this heterodyne-detected TTR technique opens up new areas in the study of molecular interactions in liquids, especially with the further development of approaches that can manipulate and detect vibrational coherences in a variety of systems -- including highly polar species.
In particular, the stage is set for a two-THz-pulse TTR experiment, where a coherence is initiated with one THz pulse and subsequently perturbed with another THz pulse, or where the time spacing between the two pulses is varied to maximize the nonlinear THz coherent excitation. This will ultimately result in a 2D TTR spectroscopy. Beyond pure liquids, there are many different condensed-phase systems such as molecular ices\cite{Allodi2014}, or polypeptides\cite{Yamaguchi2005,Yamamoto2005} that have complex THz spectra with distinct features. The dynamics of these modes, or even the specific intra- versus intermolecular nature of the modes themselves, remain poorly understood, and 2D TTR experiments would be able to address many open questions. 

The multi-cycle nature of the pulses used in this experiment would fundamentally limit the temporal resolution that could be achieved with a two-dimensional approach, but switching to a plasma-based gas photonic THz source, or thinner crystal emitters, would yield sub-cycle THz pulses -- thus opening the possibility of $<$100~fs resolution in nonlinear THz measurements.\cite{Kim2008} Such temporal resolution may be required to understand the anharmonicity of intra- and intermolecular vibrations and would provide novel data that probe how the molecular dynamics on ultrafast timescales affects the physical properties in the condensed phase. 

Extensions of this work will eventually enable coherent control of vibrational modes in liquids. Since THz pulses can directly access the soft modes that participate in the room-temperature dynamics, varying the time between pulses in a two-pulse scheme will provide one handle for the coherent excitation, as has already been demonstrated in the gas phase.\cite{Fleischer2012} In addition, controlling the shape of THz field\cite{Chen2011} can provide another handle that can be used for THz coherent control. Given the straight-forward improvements that are possible, these TTR results offer the first glimpse into unexplored realms of condensed-phase dynamics and control provided by THz radiation.

\vspace{0.5cm}

\section{Acknowledgments}
The authors gratefully acknowledge financial support from the National Science Foundation (Grants No. CHE-1214123, CHE-0722330, and the NSF Graduate Research Fellowship Program). MAA acknowledges current support from a Yen Postdoctoral Fellowship from The Institute for Biophysical Dynamics at The University of Chicago.

%

\end{document}